\documentclass[reprint,prl,twocolumn,superscriptaddress,showpacs,nofootinbib,floatfix,preprintnumbers]{revtex4-1}
\usepackage{color}              
\usepackage{graphicx, bm}
\def\lsim{\mathrel{\rlap{\lower4pt\hbox{\hskip1pt$\sim$}}
    \raise1pt\hbox{$<$}}}  
\def\gsim{\mathrel{\rlap{\lower4pt\hbox{\hskip1pt$\sim$}}
    \raise1pt\hbox{$>$}}}                

\begin{document}
\preprint{MAN/HEP/2013/15, 
CERN-PH-TH/2013-192} 
\title {\bf \Large New Production Mechanism for Heavy Neutrinos at the LHC} 
\author {P. S. Bhupal Dev}
\affiliation{Consortium for Fundamental Physics, School of Physics and
  Astronomy, University of Manchester, Manchester, M13 9PL, United
  Kingdom}  
\author{Apostolos Pilaftsis} 
\affiliation{Consortium for Fundamental Physics, School of Physics and
  Astronomy, University of Manchester, Manchester, M13 9PL, United
  Kingdom}
\affiliation{CERN, Department of Physics, Theory Division, CH-1211
  Geneva 23, Switzerland} 
\author{Un-ki Yang} 
\affiliation{Department of Physics and Astronomy, Seoul National
  University, Seoul 151-747, Korea}
\vspace{1cm}

\begin{abstract}
We study  a new production mechanism  for heavy neutrinos  at the LHC,
which dominates  over the usually  considered $s$-channel $W$-exchange
diagram for  heavy-neutrino masses larger than  100--200~GeV.  The new
mechanism   is  infrared-enhanced   by   $t$-channel  $W\gamma$-fusion
processes. This  has important implications for  experimental tests of
the seesaw  mechanism of neutrino  masses, and in particular, for the
ongoing heavy neutrino searches at the LHC.  We find that
the direct  collider limits on the light-to-heavy  neutrino mixing can
be significantly improved, when this new production
channel  is  properly taken  into  account.   The  scope of  this  new
mechanism can equally well be extended to other exotic searches at the
LHC.
\end{abstract}

\date{January 10, 2014}

\maketitle

The discovery  of nonzero neutrino  masses and mixing  from neutrino
oscillation  data provides  the  first (and  so  far only)  conclusive
experimental  evidence of  the  existence of  new  physics beyond  the
Standard Model (SM). A simple paradigm for understanding the smallness
of   neutrino    masses   in   a    natural   way   is    the   seesaw
mechanism~\cite{review1}.    Its   simplest   realization~\cite{type1}
(known as the type-I seesaw) requires  the existence of a set of heavy
SM-singlet   Majorana  fermions   $N$,   which   break  the
$(B-L)$-symmetry  of the  theory by  two  units. The  seesaw scale  is
synonymous with  the typical Majorana mass  $M_N$ of these heavy 
neutrinos,    whose   origin    must    be   connected    with   some    new
physics~\cite{review1}. In the flavor basis $\{(\nu_{L})^C, N\}$, the
seesaw  mass   matrix  has  the   following  general
structure~\cite{type1, valle}:
\begin{eqnarray}
{\cal M}_\nu\ =\ 
\left(\begin{array}{cc}0 & M_D\\ M^{\sf T}_D & M_N\end{array}\right)\;, 
\label{eq1}
\end{eqnarray}
where $M_D$ is  the Dirac mass term which mixes  the light~($\nu_L$) and
heavy ($N$)  states. 
In
the usual  seesaw approximation: $||\xi  || \ll 1$, where  $\xi \equiv
M_D M^{-1}_N$  and $||\xi || \equiv  \sqrt{{\rm Tr} (\xi^\dagger \xi)}$,
this leads to the observed light neutrino mass matrix of the form
\begin{eqnarray}
  \label{eq:1}
M_\nu\ \simeq\ -M_D M_N^{-1} M_D^{\sf T}
\end{eqnarray}
and  to the light-to-heavy neutrino  mixing of  order $\xi$~\cite{valle}.
We note  that the smallness of  $M_\nu$ could be attributed  to a very
high  value   for  $M_N$,  or   to  a  particular   flavor  structure
in~(\ref{eq:1}),  or  both.  Without  specifying  the  details of  the
model, we  generically call this minimal realization the `SM
seesaw'.

As mentioned above, there are two key aspects of the
seesaw  mechanism  that can  be  probed  experimentally: the  Majorana
mass $M_N$  of  the heavy  neutrinos,  and  the mixing  $\xi$
between the  heavy and  light neutrinos.  The  Majorana nature  of the
light and heavy neutrinos can  in principle be tested via neutrinoless    
double          beta          decay
($0\nu\beta\beta$)~\cite{0v2breview}. 
However,    this   does   not
necessarily  probe the mixing $\xi$ whose 
effects may be sub-dominant, compared to 
purely   left-(or right-)handed contributions to   the
$0\nu\beta\beta$ process.   Alternatively, a non-negligible  
value for~$\xi$ could  be inferred from  non-unitarity of  the  light
neutrino mixing matrix~\cite{unitarity}, in neutrino oscillation data,
as well as in observables for lepton flavor violation (LFV)~\cite{lfv}.  
However,
these low-energy  observables by themselves do not  prove the Majorana
nature  of heavy  neutrinos since models with pseudo-Dirac heavy  
neutrinos can also yield large non-unitarity 
and LFV effects~\cite{invunitarity}.

\begin{figure}[t]
\begin{center}
\includegraphics[width=3.2cm]{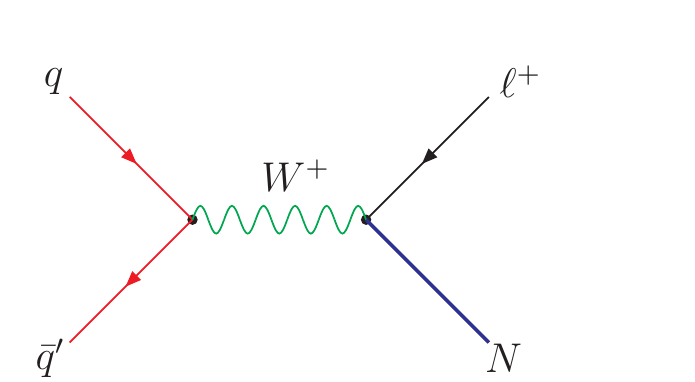}
\end{center}
\caption{The usually considered heavy neutrino production channel 
in the SM seesaw at the LHC.}
\label{fig1}
\end{figure}

In the  SM seesaw, the  Majorana nature of  possible electroweak-scale
heavy neutrinos as well as their mixing with the light neutrinos 
can  be {\it simultaneously} unraveled via their distinctive like-sign
dilepton   signatures   at   colliders~\cite{atre}.  
The  usually considered production  channel for heavy
Majorana neutrinos  at the LHC is  $pp\to W^\pm \to  \ell^\pm N$ (Fig.~\ref{fig1}), 
with $N$ subsequently  decaying to $\ell^\pm W$, followed  by the $W$-decay
to hadronic final  states.  For $M_N>M_W$, 
the  $W$-boson produced  from the  $pp$  collision is
off-shell, whereas that  coming from the $N$-decay  is on-shell. For a
Majorana  neutrino  $N$,  this  leads to  the  `smoking-gun'  collider
signature of same-sign dileptons plus  two jets with no missing energy
($\ell^\pm\ell^\pm  jj$).   This was  first  pointed  out
 in the context of Left-Right models~\cite{KS}, and        was        
subsequently        analyzed
in~\cite{APZPC,datta,theory-LL,BLP}    within     the    SM    seesaw.
Experimental searches based on this channel have been performed 
using the $\sqrt s=7$ TeV LHC data for  the di-muon case
~\cite{ATLAS-LL, CMS-LL} (and also for the  di-electron case~\cite{CMS-LL}).  
No  excess above the
expected SM background  has been observed so far,  and upper limits on
the light-to-heavy  neutrino mixing parameter  squared, $|V_{\mu N}|^2
\approx (\xi\, \xi^\dagger)_{\mu  \mu}=10^{-2}$ -- $10^{-1}$, have been
derived for heavy neutrino masses $M_N = 100$ -- 300~GeV.

For collider tests of the SM seesaw to be effective,
the  mixing  parameter  $V_{\ell  N}  \approx \xi_{\ell  N}$  must  be
significant,  since this  is  the  only way  the heavy  neutrino
communicates to  the observable SM  sector.  This requires  that apart
from  $M_N$  being   small  (in  the  sub-TeV  to   TeV  range  to  be
kinematically accessible), $M_D$ must be  large (in the few GeV range)
simultaneously.   In the traditional  ``vanilla" seesaw  mechanism, we
expect   the   light-to-heavy    neutrino   mixing   $V_{\ell   N}\sim
\sqrt{m_\nu/M_N}\lsim 10^{-7}$  for  $M_N \sim
1$~TeV, due to the smallness  of light neutrino mass $m_\nu\lsim 
0.1$    eV~\cite{Planck},   thus    making   the    collider   signal
unobservable. However, if the Dirac and Majorana mass matrices 
in (\ref{eq:1}) have specific  textures  
which  can be enforced by some symmetries~\cite{APZPC,cancel}, 
$V_{\ell N}$ can be naturally large while the light neutrinos remain 
massless at the tree-level. The observed non-zero neutrino masses and 
mixing can 
be generated by approximately breaking the underlying symmetry structure 
via radiative effects and/or higher-dimensional operators. 
Such models allow the possibility of having ${\cal O}$(100) GeV 
heavy Majorana neutrinos with a significant $V_{\ell N}$, and hence, 
observable lepton number violation (LNV)  at the LHC~\cite{ft1}, 
without being in conflict with the neutrino oscillation data. 
We will generically assume this for our subsequent  discussion, 
without referring to any particular 
texture or model-building aspects, and  so treat  $M_N$ and
$V_{\ell N}$ as free phenomenological parameters.

\begin{figure}[htb]
\begin{center}
\includegraphics[width=3.2cm]{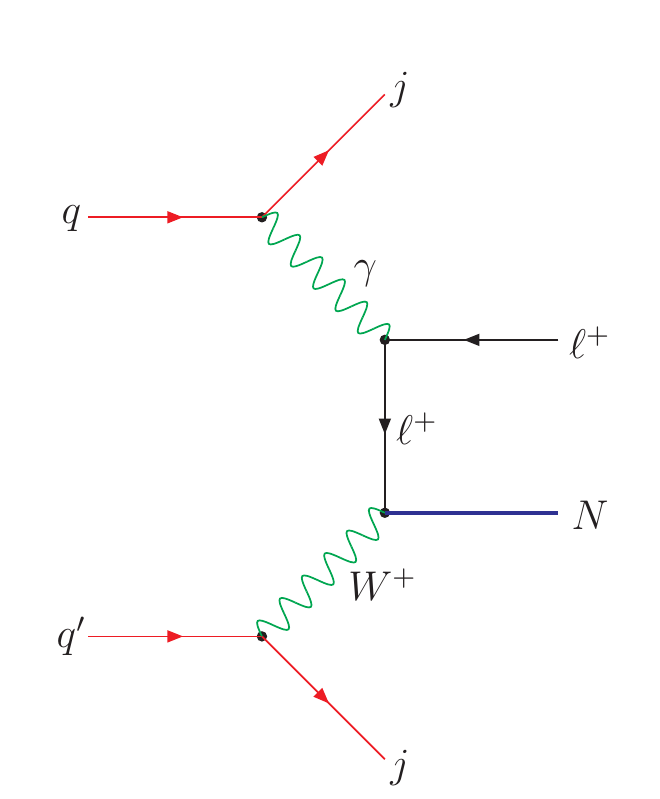}
\includegraphics[width=3.2cm]{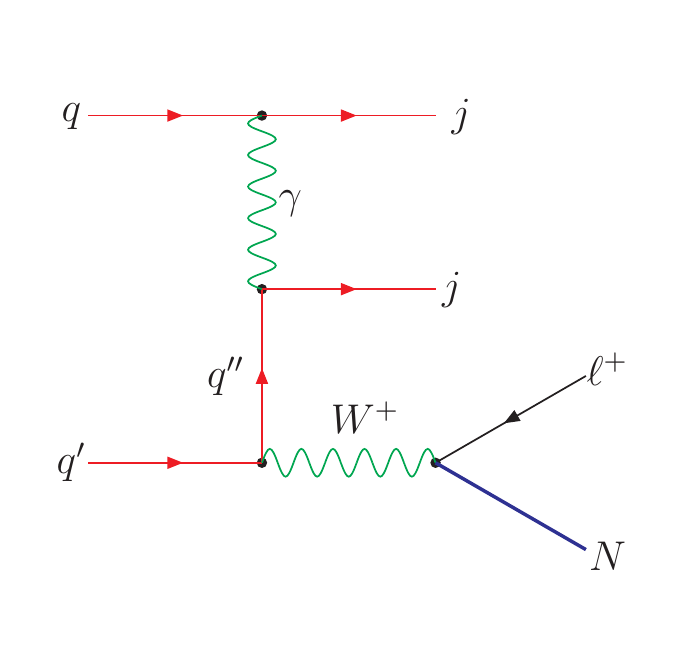}\\
\includegraphics[width=3.2cm]{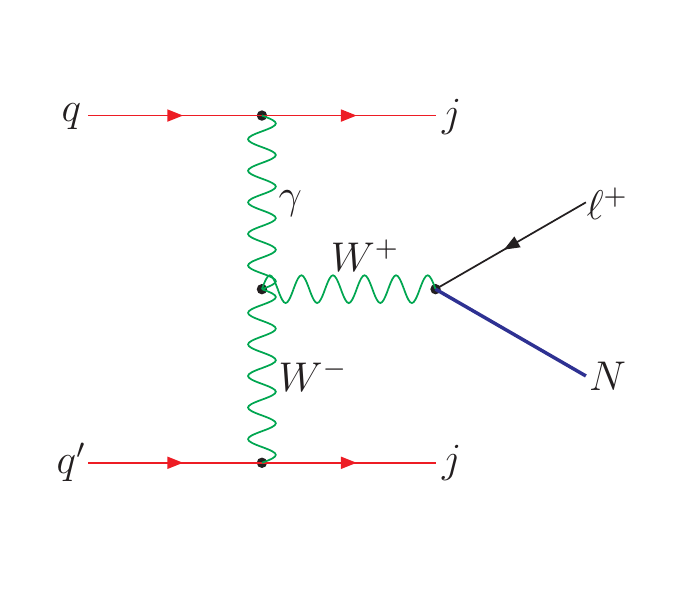}
\includegraphics[width=3.2cm]{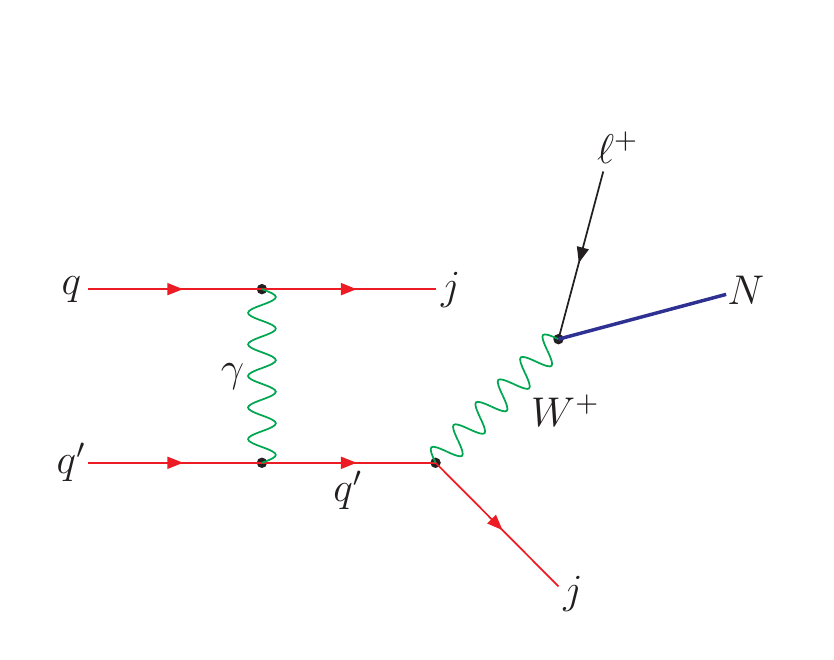}
\end{center}
\caption{New heavy neutrino production channels at the LHC.
Mirror-symmetric and $Z$-mediated graphs are not shown.} 
\label{fig2}
\end{figure}

In  this Letter  we explicitly  demonstrate the  existence of  a novel
production mechanism  for heavy neutrinos  at the LHC  which dominates
over the previously  considered $s$-channel $W$-exchange diagram shown
in Fig.~\ref{fig1}.   Within the SM seesaw,  there exist
many reactions  at parton  level listed  in~\cite{datta}, which
give  rise   to  same-sign  dileptons   with  $n\geq  2$   jets.   The
contributions  of most  of  these additional  diagrams are  negligible
compared to the that in Fig.~\ref{fig1}, and have therefore been
neglected  in  all previous  collider  analyses.   As  we show  below,
however,  diagrams involving  virtual  photons in  the $t$-channel  as
shown  in Fig.~\ref{fig2}  give rise  to {\it  diffractive} processes,
such as
\begin{eqnarray}
pp\ \to\ W^*\gamma^* jj\ \to\ \ell^\pm N jj\; ,
\label{proc2}
\end{eqnarray}
which are {\it  not} negligible, but infrared enhanced.   In fact, the
inclusive cross  section of  these processes is  divergent due  to the
collinear singularity caused by  the photon propagator. As we increase
the virtuality of the photon  by giving a large transverse momentum to
the   associated   jet    ($p_T^j$),   the   cross   section   becomes
finite. Following the Weizs\"{a}cker-Williams
equivalent photon approximation (EPA) for electrons~\cite{ww}, we may analogously write
down the  cross section as a  convolution of the  probability that the
proton radiates off a real photon, 
by absorbing the collinear divergence of
the low-$p_T^j$ regime into an effective photon structure function for
the proton~\cite{epa, epa2}.

To  establish the importance  of the  diagrams in  Fig.~\ref{fig2}, we
compare the inclusive  cross section for $N\ell^\pm jj$ 
with  the  previously  considered $N\ell^\pm$  in Fig.~\ref{fig1}.   
Note  that   the $pp\to N\ell^\pm jj$  process  receives
contributions  from  both  hadronic  and electroweak  processes.   The
hadronic channels  mediated by virtual  gluons and quarks  give ${\cal
  O}(\alpha_s)$  corrections  to   the  production  channel  in
Fig.~\ref{fig1} and  drop at  the same rate  as the  $pp\to N\ell^\pm$
cross  section,  as the  heavy  neutrino  mass  increases.   The
electroweak contributions come from the
virtual $\gamma$-exchange diagrams  shown in Fig.~\ref{fig2}, and also
from additional  $W^\pm Z$-mediated graphs not shown  here.  All these
Feynman  graphs  must  be  taken  into  account, in  order  to  get  a
gauge-invariant  result.   It turns  out  that  the total  electroweak
contribution drops at  a rate slower than the  $pp\to N\ell^\pm$ cross
section with increasing heavy neutrino mass.  This is mainly due
to  the  infrared-enhanced  cross  section  of  the  $\gamma$-mediated
processes in~\ref{proc2},  which have a significantly milder
dependence   on  $M_N$.    As   a  result,   the  production   channel
(\ref{proc2}) dominates over  the earlier considered $pp\to N\ell^\pm$
channel  with increasing  $M_N$.  Similar  behavior is  also expected
with  increasing  center  of   mass  energy  $\sqrt{s}$  in  the  $pp$
collisions,  as verified  by  our numerical  simulations given  below.
Thus,  the process~(\ref{proc2})  becomes  increasingly important  for
heavy neutrino searches at the LHC, for higher energies $\sqrt{s}$ and
also larger $M_N$ values.  Consequently, it must be taken into account
in present and future analyses of the LHC data.

\begin{figure*}[t]
\begin{center}
\includegraphics[width=5cm]{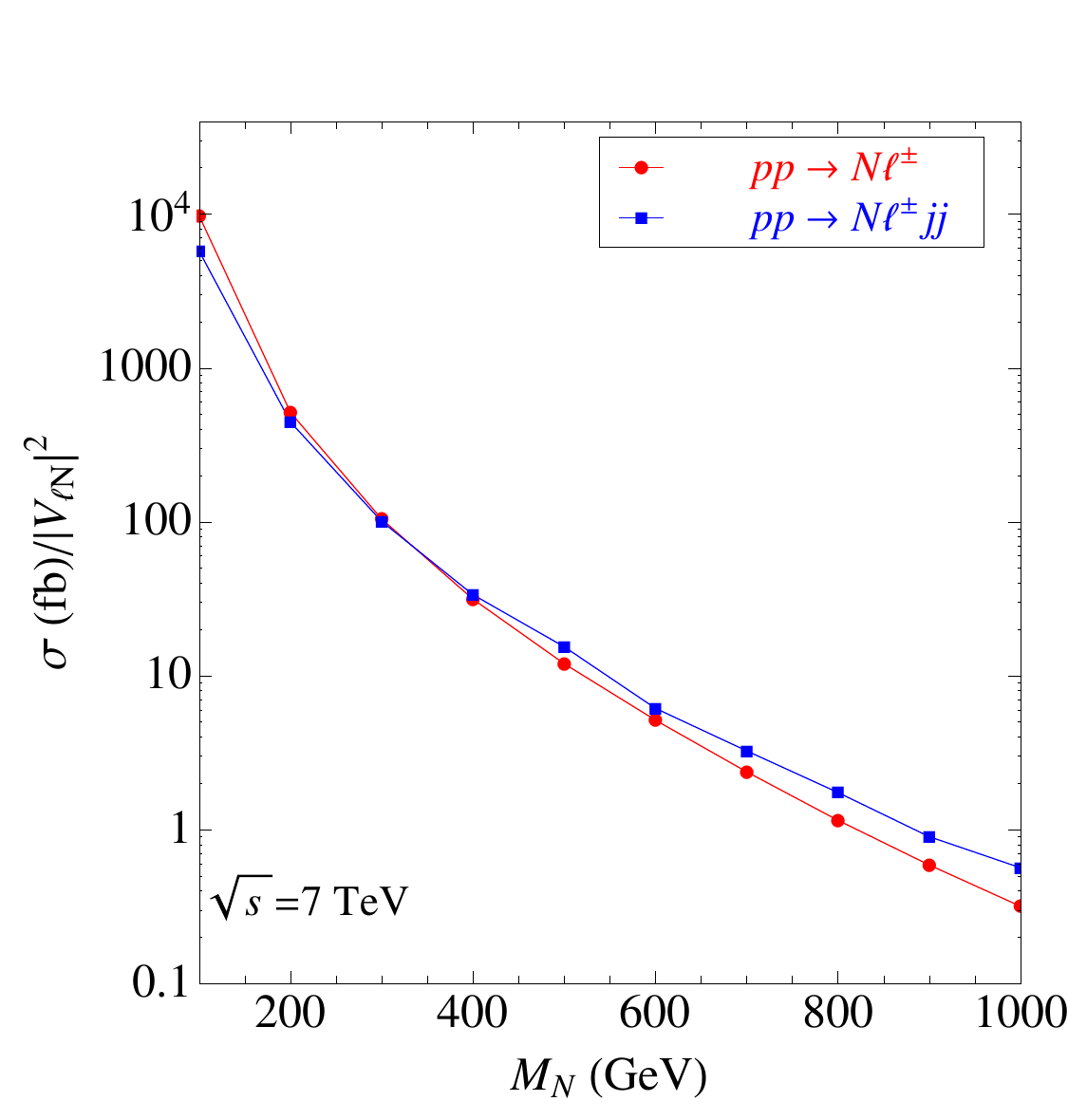}
\hspace{0.2cm}
\includegraphics[width=5cm]{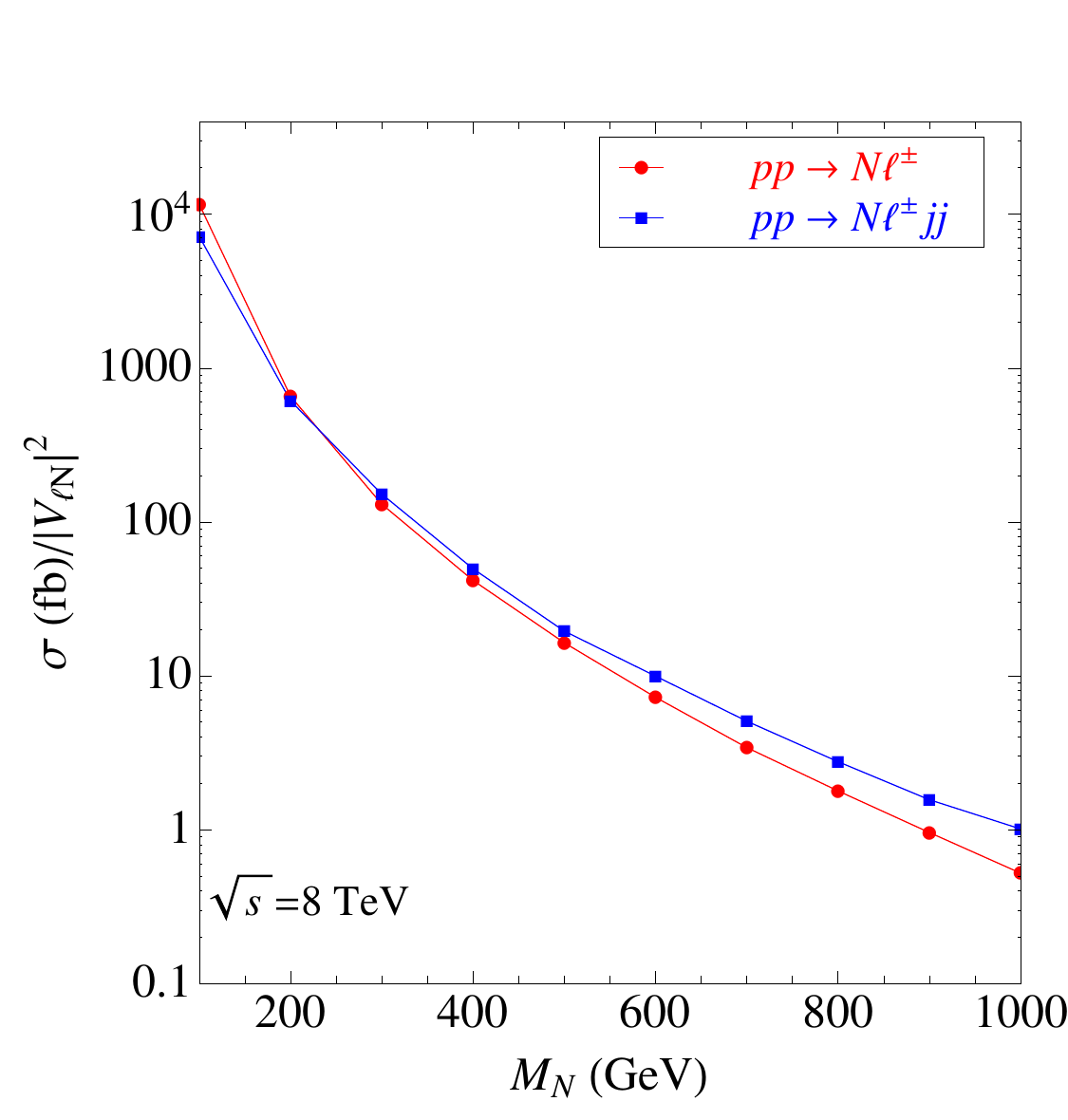}
\hspace{0.2cm}
\includegraphics[width=5cm]{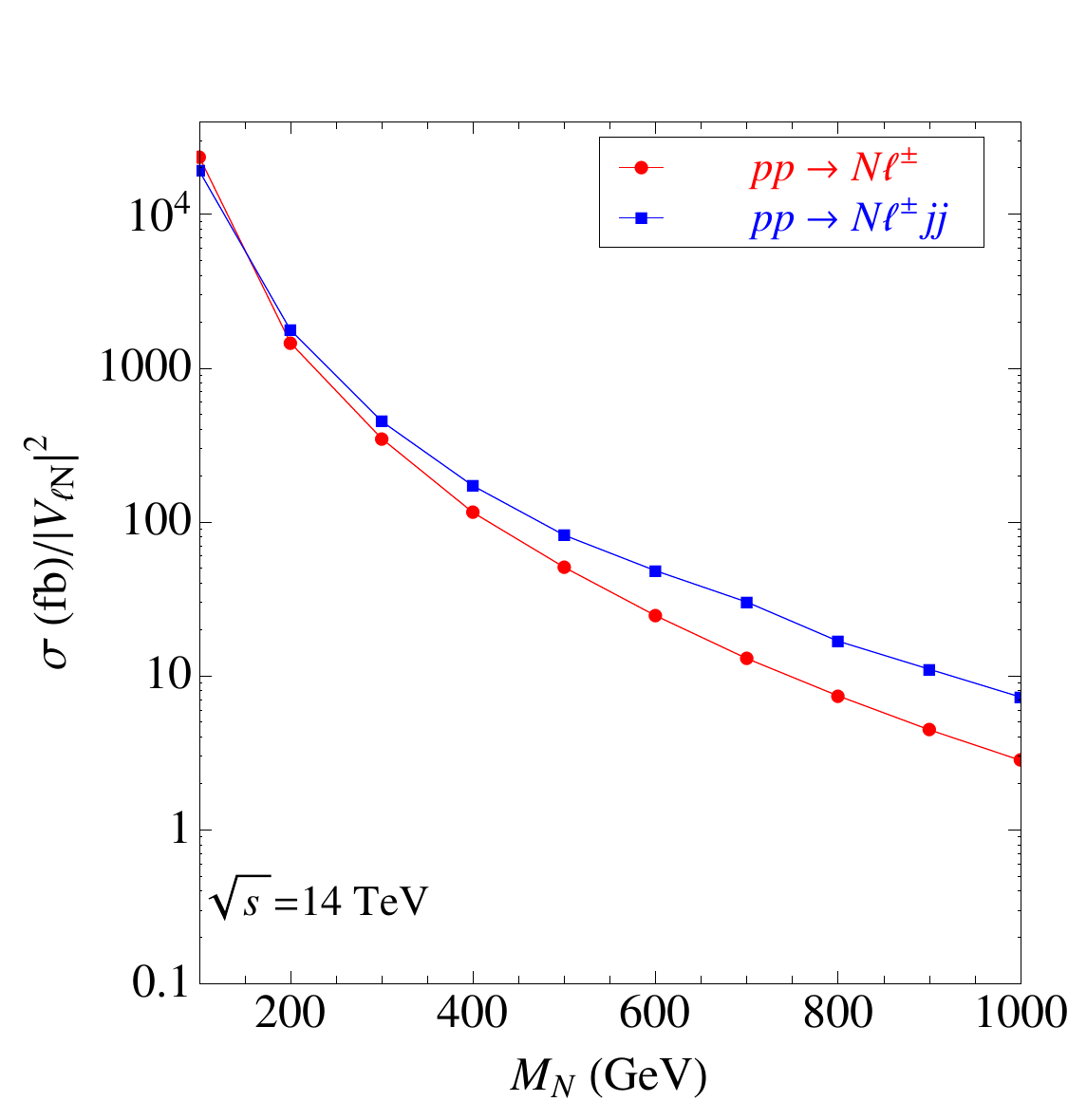}
\end{center}
\caption{Comparison of the inclusive  cross sections 
for  the heavy  neutrino production  channels $pp\to N\ell^\pm$ and $pp\to
  N\ell^\pm jj$  at LHC energies of $\sqrt s=7,8$ and 14 TeV. }
\label{fig3}
\end{figure*}
Our numerical  results are shown in Fig.~\ref{fig3}  for the inclusive
production cross sections normalized to the mixing parameter $|V_{\ell
  N}|^2=1$. For  the process $pp\to N\ell^\pm jj$, we  obtain the `inclusive'
cross   section   by   applying   a   minimal   jet   $p_T$   cut   of
$p_T^j>(p_T^j)_{\rm min}$ to  avoid the collinear singularity, whereas
the infrared  part is approximated  by the inclusive cross  section of
the process $p\gamma  \to N\ell^\pm j$, where the photon comes from a
proton.  The  latter process  was calculated with EPA 
using the improved Weizs\"{a}cker-Williams formula
~\cite{epa} for   a  fixed
factorization scale of $\mu_F=(p_T^j)_{\rm min}$. For concreteness, we
have   chosen   $(p_T^j)_{\rm  min}=10$~GeV (the lowest detection 
threshold for ATLAS) and   used  the  equivalent photon
distribution functions  as implemented in  {\tt MadGraph5}~\cite{mg5},
whereas the quark and gluon  distribution functions of the proton were
taken  from~{\tt  CTEQ6L}~\cite{cteq6l}.  The  renormalization  scale
was chosen for each event depending on the maximum final-state
mass ($M_N$ in our case). Note that the total cross section which is a
sum of  the $pp\to  N\ell^\pm jj$ and  $p\gamma\to N\ell^\pm  j$ cross
sections  should be independent  of the $p_T^j$ cut, as long  as the
collinear  part of the  $pp\to N\ell^\pm  jj$ process  is consistently
absorbed  into the  photon  distribution function. 
We observed  some
discrepancy from this  general expectation, which could be due to the fact 
that the accuracy of EPA, while being excellent for elastic scattering 
processes~\cite{kniehl}, is scale-dependent 
for inelastic channels~\cite{pisano}, and moreover, the choice of the factorization scale 
is 
not unique due to higher order effects in perturbative QCD. 
For an alternative model of EPA as currently implemented in {\tt CalcHEP3.4}~\cite{calchep}, 
we get similar results as above 
for the $p\gamma\to N\ell^\pm j$ cross sections within 10-20\% accuracy.  
However, since the
dominant contribution  results from the process  $pp\to N\ell^\pm jj$ 
in the current mass range of interest, 
this point will not affect the main results of the analysis presented here. 

From  Fig.~\ref{fig3} we  see  that the  $N\ell^\pm$  production
channel is dominant only in the  low mass regime, whilst the new
$N\ell^\pm   jj$   channel   starts   becoming   dominant   for   $M_N
\gsim 300$~GeV at $\sqrt s=7$~TeV LHC.  This crossover point  shifts towards lower values of $M_N$,
with  increasing  $\sqrt s$.   It  is  interesting  to note  that  the
existing  heavy neutrino  searches~\cite{CMS-LL,  ATLAS-LL} have  only
explored  up to  $M_N=300$~GeV  with $\sqrt  s=7$ TeV LHC  data, 
but plan to extend up to $M_N=500$ GeV
with  $\sqrt  s=8$~TeV data.  Hence,  the  new production
channel  proposed here  must be  taken into  account in
{\em all} current and future LHC analyses.

An  important consequence of  the new  production mechanism  for heavy
neutrinos is  that the current LHC sensitivity  for the light-to-heavy
neutrino mixing  parameter $V_{\mu  N}$ can be  improved significantly
for the whole heavy neutrino  mass range of interest, i.e.~$M_N = 100$
- 300~GeV. In order to derive the  new limits on $V_{\mu N}$, we first
calculate  the efficiency  of  the new  signal  proposed here:  $pp\to
N\mu^\pm  jj\to  \mu^\pm  \mu^\pm  4j$, after  implementing  the  same
selection criteria as used for  the $\mu^\pm\mu^\pm jj$ channel in the
$\sqrt s=7$ TeV ATLAS analysis~\cite{ATLAS-LL}:
\begin{eqnarray}
&&p_T^j>20~{\rm GeV},~p_T^\mu >20~{\rm GeV},~
p_T^{\mu,{\rm leading}}>25~{\rm GeV},\nonumber\\
&& |\eta^j|<2.8,~|\eta^\mu|<2.5, ~\Delta R^{jj}>0.4, ~\Delta R^{\mu
  j} > 0.4, \label{select}\\ 
&& m_{\mu\mu} > 15~{\rm GeV}, ~E_T^{\rm miss}<35~{\rm GeV}, ~m_{jj} \in
[55,120]~{\rm GeV}, \nonumber 
\end{eqnarray}
and for $\Delta R^{\mu  j}<0.4$, we require  $p_T^\mu>80$~GeV to
retain muons  close to jets  from event topologies with  boosted heavy
neutrinos.   After  generating  the  parton  level  events  with  {\tt
  MadGraph5}~\cite{mg5},   the   showering   and  hadronization   were
performed  with  {\tt  Pythia6.4}~\cite{pythia}  and a  fast  detector
simulation was done using {\tt DELPHES2.0.5}~\cite{delphes}.  Jets are
reconstructed  using the  anti-$k_T$ jet  clustering algorithm  with 
$R=0.4$ as  implemented in {\tt FastJet2}~\cite{fastjet}. We
find  that the  total  selection efficiency  for the  $\mu^\pm\mu^\pm$
signal remains  almost the  same as before~\cite{ATLAS-LL},  since the
additional two jets  coming from the new channel  are usually lost due
to the  stringent selection criteria given in  (\ref{select}).  Regarding 
the  SM background  for these  processes,  we expect the background for
di-muon+$n$  jets (with $n\geq  2$) to  be the  same as  that reported
in~\cite{ATLAS-LL} for the  selection criteria in (\ref{select}). 
Note that the SM backgrounds for the $\mu^\pm\mu^\pm 4j$ 
signal reported here mainly come from $t\bar t+V$ (where $V=W,Z$) and $WW$ production, which are small compared to the $WZ$ background 
for the $\mu^\pm\mu^\pm jj$ signal~\cite{ATLAS-LL}. 
A separate dedicated set of selection criteria and
background reduction methods must  be designed in order to distinguish
the new $\mu^\pm\mu^\pm 4j$  signal from the usual $\mu^\pm\mu^\pm jj$
signal,  and this  will be  studied  elsewhere.  A  similar
analysis can  also be performed  for the di-electron signal $e^\pm  e^\pm nj$ 
(with $n\geq 2$). Although  the   limits  on  $|V_{eN}|^2$   derived  from
$0\nu\beta\beta$    constraints   are    much   more
stringent~\cite{atre}, models  with  quasi-degenerate heavy
Majorana neutrinos  may naturally evade  these constraints, while 
giving rise to sizable LNV signals at the LHC~\cite{BLP}.
For  the  corresponding  limits  on  $|V_{\tau  N}|^2$,  the
identification  of  same-sign  di-tau  events  at  the  LHC  is  quite
difficult, thus making a realistic collider simulation for this case rather involved.

\begin{figure}[t]
\begin{center}
\includegraphics[width=8cm]{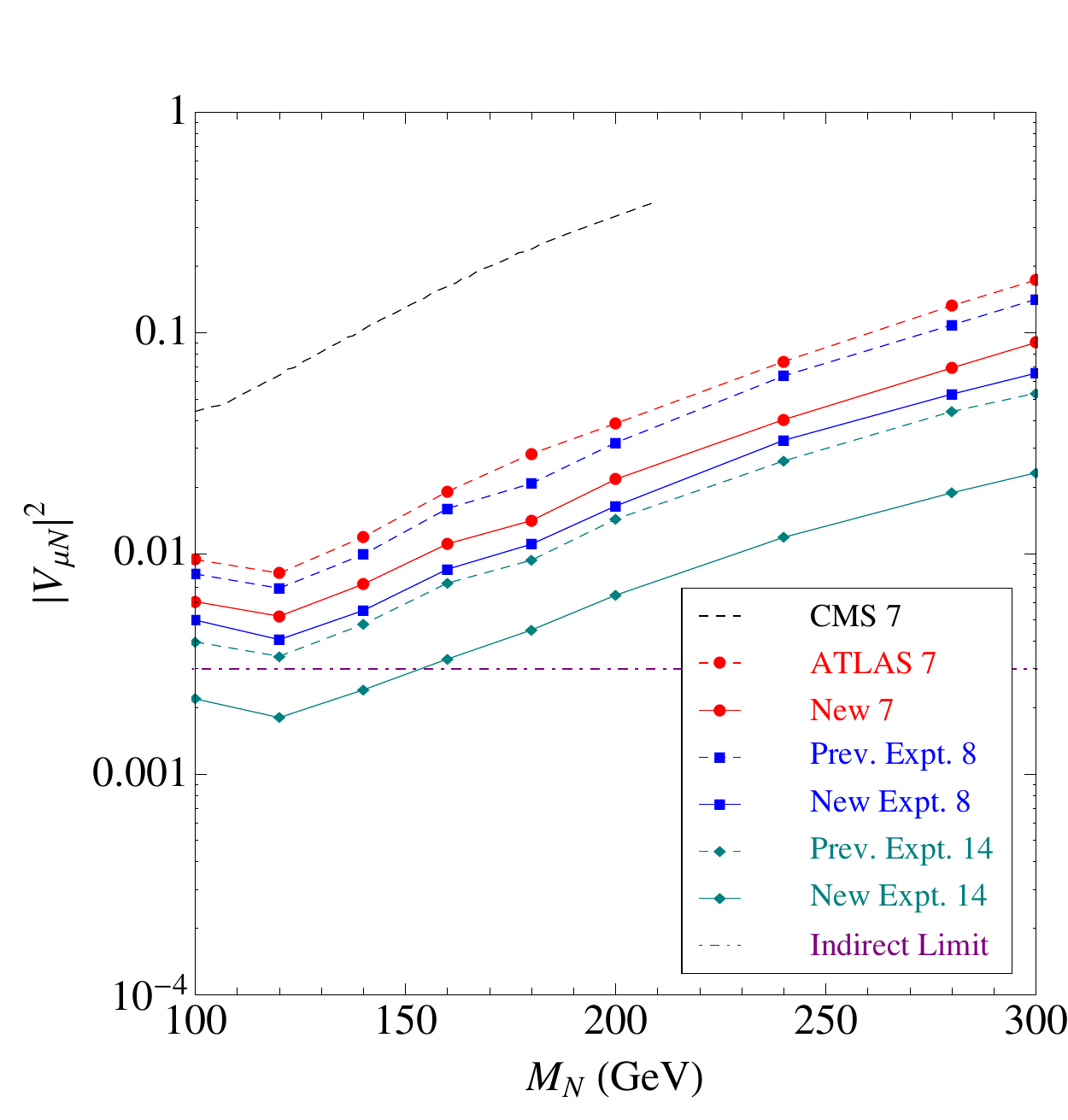}
\end{center}
\caption{Improved upper  limits (solid lines, also labeled ``New") on the  mixing
  parameter $|V_{\mu  N}|^2$ for
  LHC  energies $\sqrt  s=7, 8$  and  14~TeV, along  with the  current
  CMS~\cite{CMS-LL}   and  ATLAS~\cite{ATLAS-LL}  limits   derived  at
  $\sqrt{s} =  7$~TeV, and the conservative upper limits expected for 
  8-  and  14-TeV LHC  runs using the  {\em  previous}
  production mode  of Fig.~\ref{fig1} (dashed lines, also labeled ``Prev.''). 
The  horizontal
  line  shows the  current best limit  on $|V_{\mu  N}|^2$ from
  indirect searches~\cite{del}.}
\label{fig4}
\end{figure}

Following a rather conservative approach  to our analysis here, we use
the  current 95\%   confidence  level  upper  limits  on   the  cross  section
$\sigma(pp\to  \mu^\pm\mu^\pm jj)$~\cite{ATLAS-LL}, derived from the $\sqrt  s=7$  TeV  LHC data 
with 4.7 fb$^{-1}$ luminosity, and  translate them
into upper limits on the  mixing parameter $|V_{\mu N}|^2$ as shown in Fig.~\ref{fig4} 
by dividing the cross section limits by the total inclusive cross section 
$\sigma(pp\to \mu^\pm\mu^\pm nj)$ (with $n\geq 2$).   
We  find that  the  existing  ATLAS limits~\cite{ATLAS-LL} are  
improved  by  almost 50\%  with  the
inclusion of  the new production  mechanism.  For comparison,  we also
show the corresponding CMS  limits~\cite{CMS-LL} which are much weaker
compared to  those by  ATLAS, mainly due  to their  large backgrounds.
The  horizontal  line  shows  the  current  best limit  on
$|V_{\mu   N}|^2$  derived   indirectly  from   electroweak  precision
data~\cite{del} which is independent of the heavy neutrino mass
for $M_N>M_Z$. Note that the LFV processes (such as rare lepton 
decays~\cite{IP} and $\mu-e$ conversion~\cite{conv}) put stringent constraints   
on the product $|V_{\ell N}V^*_{\ell'N}|$ (with $\ell\neq \ell'$)~\cite{lfv}, thereby 
limiting the LHC sensitivity for LFV signals of the type $e^\pm\mu^\pm jj$; however, 
they do not restrict the individual mixing parameters $|V_{\ell N}|^2$. 
In  order to compare the direct  search limits with the
indirect one, we also derive our expected upper limits for $\sqrt s=8$
and 14 TeV  LHC by assuming that the  corresponding experimental upper
limits on  the signal cross  section will be  {\it at least} as good  as the
$\sqrt s=7$ TeV results.  Again,  these are conservative limits as the
experimental  limits   on  cross  section  are   expected  to  improve
significantly  with  the  analysis  of  more data,  if  no  signal  is
observed. In that case, the direct collider limits  could 
surpass the indirect limits for a  significant range of heavy neutrino masses,
once  the new production  mechanism proposed  here is  considered.  In
particular,  Fig.~\ref{fig4}   shows  that  the  effect   of  the  new
production mechanism  at LHC energies  $\sqrt{s} = 14$~TeV will  be to
improve the current ATLAS limit by {\it at least} a factor of five.

In summary, we  have analyzed a new dominant  production mechanism for
heavy neutrinos at the LHC.  This mechanism is extremely important for
the range of  heavy neutrino masses currently being searched
for  and provides significantly  improved {\em  direct} limits  on the
light-to-heavy neutrino  mixing $V_{\ell  N}$, in a  fully independent
fashion of the indirect searches. As more data are gathered at the LHC
and  the   sensitivity  to  higher  heavy  neutrino   mass  ranges  is
contemplated, these  new contributions will be crucial in
setting  the  best possible  direct  limits  on  the mixing  parameter
$V_{\mu  N}$ in the absence of a signal. On the other hand, 
an evidence of LNV at the LHC could   
reveal underlying symmetries of the lepton sector, thus  
shedding light on the  seesaw mechanism. 
We should note that the scope of the new infrared-enhanced 
production mechanism
proposed here is not just limited to heavy Majorana neutrinos, and can
also  be  applied to  other  heavy  particle  searches (the  so-called
`exotics')  at  the  LHC. For instance, for pseudo-Dirac heavy neutrinos, 
 the same production channels studied 
here could give rise to an enhanced tri-lepton signal. 
This mechanism is also applicable for searches of 
vector-like fermions and new charged  scalars. 
We hope to  address some of these
aspects in a future communication.

\acknowledgments 
P.S.B.D. and A.P. thank Mike Seymour for helpful discussions on the
equivalent photon approximation. P.S.B.D. also thanks Francisco del Aguila 
and Rabindra Mohapatra for useful discussions on lepton number violation, 
Olivier Mattelaer for
help with photon structure functions in {\tt MadGraph}, 
John Almond and Joel Klinger for
clarifications on the ATLAS selection cuts used here. The work of
P.S.B.D. and A.P. is supported by the Lancaster-Manchester-Sheffield
Consortium for Fundamental Physics under STFC grant ST/J000418/1. In
addition, A.P.  gratefully acknowledges partial support by a IPPP
associateship from Durham University.


\begin{thebibliography}{99}
\bibitem{ft1} Even for quasi-degenerate heavy Majorana neutrinos (as occurs in specific models), the LNV signal can be sizable when the mass splitting $\Delta M_N\equiv M_{N_1}-M_{N_2}$ is comparable to the width $\Gamma_N\equiv (\Gamma_{N_1}+\Gamma_{N_2})/2$~\cite{BLP}. For instance, for the on-shell production of $N_{1,2}$ with average four-momentum squared $\bar{s}=(M_{N_1}^2+M_{N_2}^2)/2$, the LNV amplitude in the muon-channel is given by ${\cal A}_{\rm LNV}^{\mu\mu}(\bar{s})=-V^2_{\mu N_1}\frac{2\Delta M_N}{\Delta M_N^2+\Gamma_N^2}+{\cal O}\left(\frac{\Delta M_N}{M_{N_1}}\right)$, for $\Delta M_N\lsim \Gamma_N$.
\end{thebibliography}
\end{document}